\documentclass[prb,reprint,superscriptaddress]{revtex4-1}
\usepackage[caption=false]{subfig}
\usepackage[version=4]{mhchem}
\usepackage{graphicx,color}
\usepackage[usenames,dvipsnames,svgnaes,table]{xcolor}
\usepackage{bm}
\usepackage{amsmath}
\usepackage{amssymb}
\usepackage{mathtools}
\usepackage{siunitx}
\usepackage{multirow}
\usepackage{array}
\usepackage{color}
\usepackage{longtable}
\usepackage{xfrac}
\usepackage{hyperref}
\hypersetup{colorlinks=true,
            linkcolor=red,
            citecolor=blue,
            urlcolor=orange}
\usepackage{graphicx, type1cm, lettrine}
\newcolumntype{K}[1]{>{\centering\arraybackslash}p{#1}}
\setcitestyle{numbers,round}

\newif\ifdraft
\drafttrue

\ifdraft
\newcommand{\amsler}[1]{{\textcolor{red}{ Max: #1 }}}
\newcommand{\vinay}[1]{{\textcolor{green}{ Vinay: #1 }}}
\newcommand{\logan}[1]{{\textcolor{teal}{ Logan: #1 }}}
\else
\newcommand{\amsler}[1]{}
\newcommand{\vinay}[1]{}
\newcommand{\logan}[1]{}
\fi

\AtBeginDocument{\usepackage{booktabs}}

\begin{document}

\title{Thermodynamics and superconductivity of S$_x$Se$_{1-x}$H$_3$}

\author{Maximilian Amsler}
\email{amsler.max@gmail.com}
\affiliation{Laboratory of Atomic and Solid State Physics,
Cornell University, Ithaca, New York 14853, USA}
\date{\today}

\begin{abstract}
The compression of \ce{SH2} and its subsequent decomposition to \ce{SH3}, presumably in a cubic \textit{Im$\overline{3}$m} structure, has lead to the discovery of conventional superconductivity with the highest measured and confirmed $T_c$ to date, 203~K at 160~GPa. Recent theoretical studies suggest that a mixture of S with other  elements of the chalcogen group could improve the superconducting temperature. Here, we present a detailed analysis of the thermodynamic properties of S and Se mixtures in the  bcc  lattice with \textit{Im$\overline{3}$m}  symmetry using a cluster expansion technique to explore the phase diagram of \ce{S_xSe_{1-x}H_{3}}. In contrast to earlier reports, we find that  \ce{S_{0.5}Se_{0.5}H3} is not stable in the  pressure range between 150--200~GPa. However, phases at compositions \ce{S_{0.2}Se_{0.8}H3}, \ce{S_{0.$\overline{3}$}Se_{0.$\overline{6}$}H3}, and \ce{S_{0.6}Se_{0.4}H3} are stable at 200~GPa, while additional phases at \ce{S_{0.25}Se_{0.75}H3} and \ce{S_{0.75}Se_{0.25}H3} are accessible at lower pressures. Electron-phonon  calculations show that the values of $T_c$ are consistently lower for all ternary phases, indicating that mixtures of S and Se with H might not be a viable route towards compounds with improved superconducting properties.

\end{abstract}
\maketitle

Metallic hydrogen~\cite{wigner_possibility_1935} has become the holy grail in high-pressure  physics due to its predicted exotic properties, most notably the expected high-$T_c$ superconductivity in its molecular or atomic form, potentially above room temperature~\cite{ashcroft_metallic_1968,cudazzo_ab_2008,mcmahon_high-temperature_2011,borinaga_anharmonic_2016}. Despite recent reports on the successful formation of metallic hydrogen in diamond anvil cells at static pressures close to 500~GPa~\cite{dias_observation_2017}, the findings still remain unconfirmed and are subject to controversial discussions~\cite{goncharov_comment_2017,silvera_response_2017,eremets_comments_2017,loubeyre_comment_2017,liu_comment_2017}. On the other hand, Ashcroft's proposal~\cite{ashcroft_hydrogen_2004} to lower the metalization pressure of hydrogen by adding heavier elements to exert chemical pressure in hydrogen-rich compounds has proven to be particularly fruitful. Theoretical studies based on structural searches and \textit{ab initio} calculations have been performed to screen for many potential candidate materials, ranging from silicon~\cite{kim_crystal_2008,martinez-canales_novel_2009,flores-livas_high-pressure_2012}, scandium~\cite{ye_high_2018}, sulfur~\cite{li_metallization_2014,duan_pressure-induced_2014}, and phosphorus hydrides~\cite{shamp_decomposition_2016,liu_crystal_2016,flores_Superconductivity_2016} to calcium, lanthanum and yttrium hydrides~\cite{wang_superconductive_2012,liu_potential_2017,peng_hydrogen_2017} with very high hydrogen content.

The existence of high-$T_c$ hydride compounds has been meanwhile reported in at least three chemical systems through high-pressure experiments, namely in \ce{PH_x}~\cite{drozdov_superconductivity_2015}, \ce{LaH_x}~\cite{drozdov_superconductivity_2018,somayazulu_evidence_2018},  and \ce{SH_x}~\cite{drozdov_conventional_2015,troyan_observation_2016,einaga_crystal_2016}. For the latter, compression of \ce{SH2} up to 250~GPa~\cite{drozdov_conventional_2015} has lead to the discovery of two distinct regimes of superconductivity, namely a low-$T_c$ phase  (33-150~K) and a high-$T_c$ (203~K) phase, the highest measured and confirmed superconducting transition temperature to date. The different superconducting states emerge depending on the synthesis conditions, and the common consensus is that the high-$T_c$ phase can be attributed to a decomposition of \ce{SH2} to \ce{SH3} in annealed samples~\cite{bernstein_what_2015,errea_high-pressure_2015,errea_quantum_2016,li_dissociation_2016,ishikawa_superconducting_2016,flores-livas_high_2016,einaga_crystal_2016,akashi_possible_2016,kruglov_refined_2017,yao_superconducting_2018}. According to crystal structure prediction (CSP) calculations~\cite{duan_pressure-induced_2014,flores-livas_high_2016,kruglov_refined_2017} and in agreement with available experimental data~\cite{einaga_crystal_2016}, the structure of this  \ce{SH3} phase has a bcc lattice with \textit{Im$\overline{3}$m} symmetry, and is stable at pressures above about 150~GPa. The values of $T_c$ predicted from Eliashberg theory are very close to the experimental measurements~\cite{duan_pressure-induced_2014,errea_high-pressure_2015,flores-livas_high_2016}. This excellent agreement between theory and experiment together with the isotope effect measurements~\cite{drozdov_conventional_2015} and the recent optical spectroscopy studies~\cite{capitani_spectroscopic_2017} confirm that \ce{SH3} is indeed a conventional, phonon-mediated superconductor.

The success of \textit{ab initio} calculations to accurately describe the fascinating properties of \ce{SH3} has turned this system into a playground to test new ideas that could further enhance its properties. Heil~\textit{et al.}~\cite{heil_influence_2015} replaced the S atoms with chalcogens (O, S, Se, Te) using the virtual crystal approximation in an attempt to identify trends that would increase the $T_c$, and found that a partial  substitution of S with O could enhance its value. Ge~\textit{et al.}~\cite{ge_first-principles_2016} proposed doping \ce{SH3} with elements from neighboring groups in the periodic table, and concluded that a $T_c$ as high as 280~K could be reached at 250~GPa in \ce{S_{0.925}P_{0.075}H3}. Very recently, Liu~\textit{et al.}~\cite{liu_effect_2018} performed CSP calculations at a fixed composition of \ce{S_{0.5}Se_{0.5}H_{3}} and found that the lowest enthalpy structures indeed correspond to different decorations of the cubic \ce{SH3} lattice. Based on their electron-phonon  calculations, the superconducting temperature decreases when S is replaced by Se, which the authors attribute to a decreasing strength of the covalent H--S or H--Se bonds.

In this work, we investigate the thermodynamic and superconducting properties of the complete compositional range of \ce{S_xSe_{1-x}H_{3}}. Using a cluster expansion (CE) of the cubic lattice of \ce{SH3}, we sample all phases with up to 56 atoms/cell at pressures between 150 and 200~GPa. In contrast to earlier reports, we discover that \ce{S_{0.5}Se_{0.5}H_{3}} is not thermodynamically stable at any pressure. However, phases with compositions \ce{S_{0.2}Se_{0.8}H3}, \ce{S_{0.$\overline{3}$}Se_{0.$\overline{6}$}H3}, and \ce{S_{0.6}Se_{0.4}H3} are stable at 200~GPa, while \ce{S_{0.25}Se_{0.75}H3} and \ce{S_{0.75}Se_{0.25}H3} are stable at lower pressures. Our calculations reveal that all phases are superconductors, but exhibit $T_c$ values significantly lower than  pure  \ce{SH3} and  \ce{SeH3}.

\begin{figure}[htb!]
	\centering
	\includegraphics[width=0.5\columnwidth]{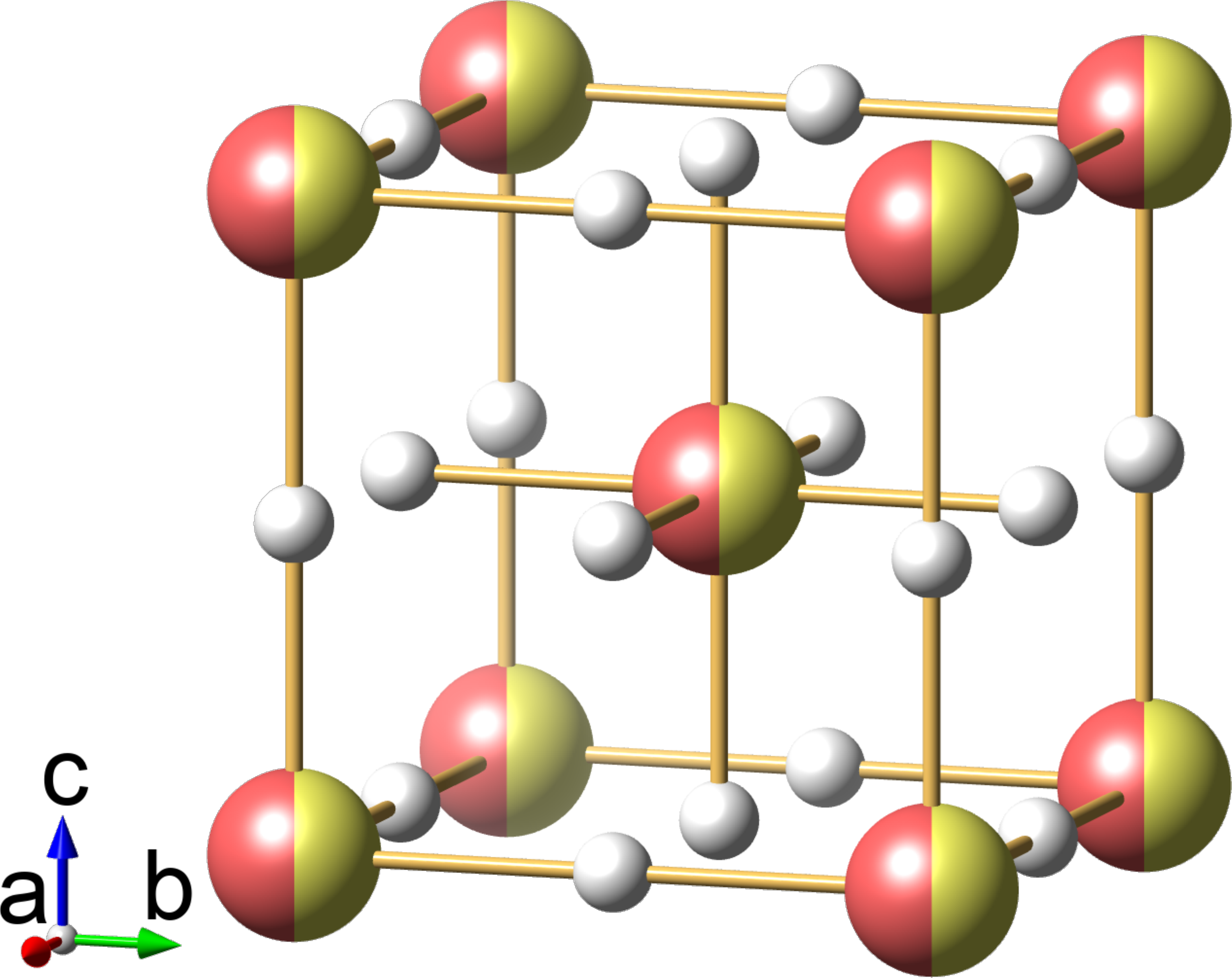}
	\caption{The parent structure of \ce{S_xSe_{1-x}H_{3}} with cubic \textit{Im$\overline{3}$m} symmetry. The small (white) spheres denote the H sites which are fully occupied. The large spheres (orange and yellow) denote the mixing sites with variable occupation of either  S or Se.}\label{fig:Structure}
\end{figure}

We start out by showing the conventional unit cell of the high-pressure phase of \ce{SH3} and \ce{SeH3} with \textit{Im$\overline{3}$m} symmetry in Fig.~\ref{fig:Structure}, where the large spheres denote the S and Se sites, and the small spheres represent the H atoms. Recent theoretical studies report that the phase diagram of Se--H  is similar to S--H, and both systems crystallize in this particular structure: The phase transitions from low-pressure phases occur above 150 and 100~GPa for \ce{SH3} and \ce{SeH3}, respectively~\cite{duan_pressure-induced_2014,flores-livas_high_2016,kruglov_refined_2017}. The S/Se atoms form a body centered cube, while the H atoms are centered between neighboring S/Se atoms. This atomic arrangements leads to two interpenetrating cubic lattices where the edges are formed through linear S/Se--H--S/Se units. Density functional theory (DFT) calculations have shown that the chemical bonds are predominantly covalent, giving rise to strong electron-phonon interactions that ultimately lead to the record-high superconducting temperature in \ce{SH3}~\cite{heil_influence_2015}.

\begin{figure}[htb!]
	\centering
	\includegraphics[width=1\columnwidth]{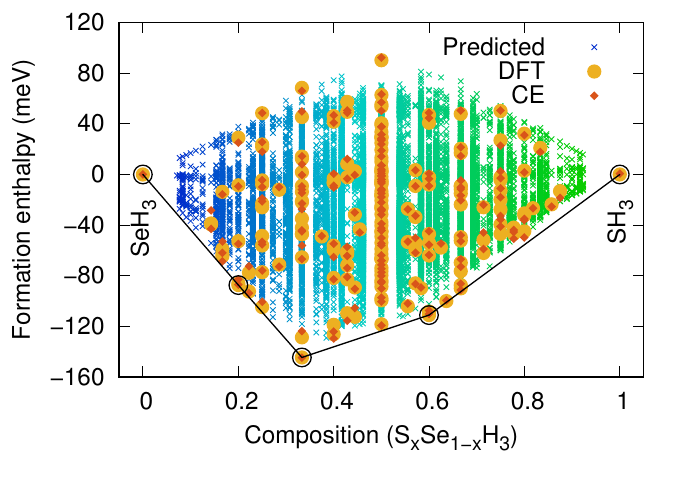}
	\caption{
	The formation enthalpies (in units of meV/\ce{XH3}) and the corresponding convex hull of the \ce{S_xSe_{1-x}H_{3}} system at 200~GPa. The end points are the binary phases \ce{SeH3} and \ce{SH3} in their respective \textit{Im$\overline{3}$m} ground state structures. The crosses denote structures that are predicted from the CE. The yellow filled circles correspond to cluster configurations that are computed with DFT (and used for the ECI fit), while the red squares represent the predicted enthalpies from the corresponding CE.}\label{fig:Hull200GPa}
\end{figure}

The central motivation in the recent work of Liu~\textit{et al.}~\cite{liu_effect_2018} was to address the issue that so far all studies in the S--Se--H system have been only treated within the virtual crystal approximation, without explicitly taking into account any potential changes in the underlying crystal structure. To this end, the authors performed CSP calculations at a fixed composition of \ce{S_{0.5}Se_{0.5}H_{3}}. Their results showed that an ordered structure was preferred over structural disorder. However, all low-enthalpy phases that they found during their structural search are merely different decorations of the S/Se sites in the   \textit{Im$\overline{3}$m}    parent lattice.

These findings raise the question if other decorations of the lattice with different compositions might have lower formation enthalpies. To address this issue, we use the  cluster expansion technique~\cite{sanchez_generalized_1984,fontaine_cluster_1994},  which is frequently employed to study metals and alloys, and provides a convenient means to expand the enthalpy in terms of short-range structural arrangements. We use  the Alloy-Theoretic Automated Toolkit (ATAT)~\cite{avdw:maps,avdw:atat,avdw:atat2} to perform a CE based on the formation enthalpies from first principles DFT calculations. In a CE, sites $i$ in a lattice  are assigned an occupation variable $\sigma_i$, depending on the atom type. A specific arrangement of these $\sigma_i$, called a configuration, is encoded in a vector $\sigma$, and the energy (or enthalpy) of said configuration is expressed in terms of ``clusters'' $\alpha$  through 
\begin{equation}
	E(\sigma)= \sum_\alpha m_\alpha J_\alpha \left< \prod_{i\in \alpha '} \sigma_i\right>
	\label{eq:CE}
\end{equation}
$\alpha$ represents a set of sites $i$ that are symmetrically inequivalent, and for every $\alpha$ we take the average over all clusters $\alpha '$ which are symmetrically equivalent to $\alpha$ with multiplicity  $m_\alpha$. The effective cluster interactions (ECI) $J_\alpha$ are fitted from a rather small set of configurations and their DFT enthalpies. In this way, the enthalpy of any configuration $\sigma$ can be quickly evaluated through equation~\eqref{eq:CE}, allowing a fast exploration of the enthalpy as a function of compositions. Here, we use occupational variation on the S/Se sites of the \textit{Im$\overline{3}$m} lattice, keeping the H atoms fixed and fully occupied.

The DFT calculations to fit the ECI are performed with the Vienna Ab initio Simulation Package (VASP)~\cite{kresse1993ab, kresse1996efficiency,kresse1996efficient} within the projector augmented wave (PAW) formalism~\cite{blochl1994projector, kresse_paw_1999}, using the PBE parameterization of the generalized gradient approximation to the exchange correlation functional~\cite{perdew1996generalized}. For the CE, we use $k$-point meshes with about  8000~$k$-points per reciprocal atom together with a plane-wave cutoff energy of 500~eV. The structural relaxations are carried out by taking into account the atomic and cell degrees of freedom until the force components on the atoms are within 0.01~eV/\AA{}, and stresses are within a few kbar. For phases that are predicted to be the ground states from the CE, we refine the enthalpies by performing iterative variable cell shape relaxations until the forces are smaller than 0.002~eV/\AA{}.

Fig.~\ref{fig:Hull200GPa} shows the results of our CE of the  \ce{S_xSe_{1-x}H_{3}} system at 200~GPa. Some 170 configurations are used to fit the ECI, giving rise to a very accurate cross-validation score of 10~meV/site. The filled yellow circles denote the configurations that are evaluated with DFT calculations, while the crosses and squares correspond to the predicted enthalpies from the CE. The convex hull construction shows that configurations at the compositions \ce{S_{0.2}Se_{0.8}H3}, \ce{S_{0.$\overline{3}$}Se_{0.$\overline{6}$}H3}, and \ce{S_{0.6}Se_{0.4}H3} are thermodynamically stable, but not at the composition \ce{S_{0.5}Se_{0.5}H_{3}} explored by Liu~\textit{et al.}~\cite{liu_effect_2018} (see Supplementary Materials for all ground state structures). However, we find that the lowest energy structure at \ce{S_{0.5}Se_{0.5}H_{3}} corresponds to the putative ground state determined through the CSP exploration of Liu~\textit{et al.} with \textit{Fd$\overline{3}$m} symmetry, providing additional confidence that the cluster expansion is  well converged.

Further CE calculations at 175 and 150~GPa show that additional phases become thermodynamically stable at lower pressures (see Supplementary Materials). In particular, the compositions \ce{S_{0.25}Se_{0.75}H3} and \ce{S_{0.75}Se_{0.25}H3} are stable at 150~GPa. However, at no pressure does \ce{S_{0.5}Se_{0.5}H3} touch the convex hull of stability. Since DFT calculations have shown that a rhombohedral \textit{R3m} phase  of \ce{SH3} becomes stable below 150~GPa~\cite{flores-livas_high_2016}, the CE results for the \textit{Im$\overline{3}$m}  parent lattice might not be representative at these pressures. Therefore, all further discussions will be restricted to calculations at 200~GPa.

\begin{figure}[htb!]
	\centering
	\includegraphics[width=0.8\columnwidth]{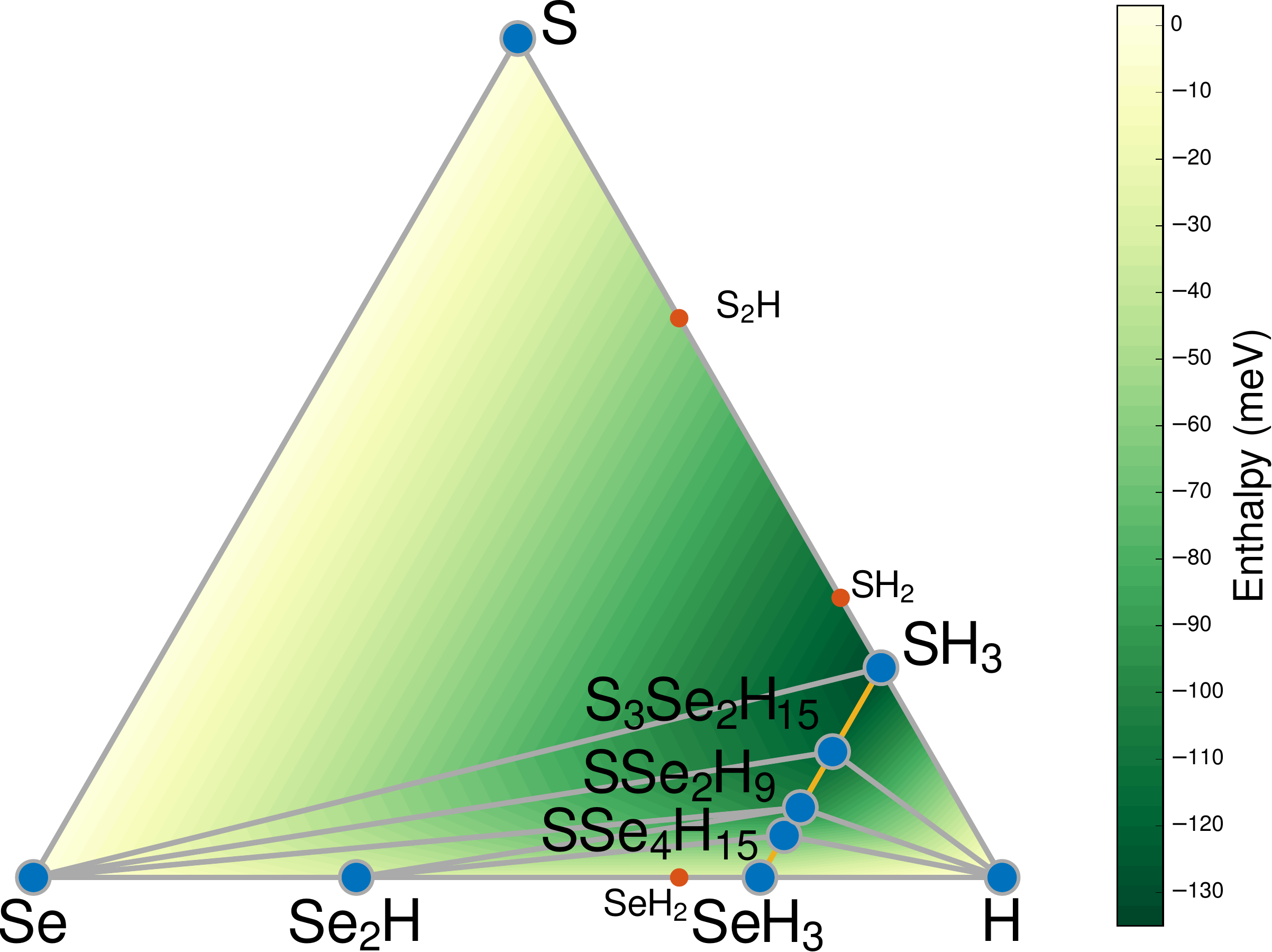}
	\caption{The Gibbs triangle convex hull of the ternary phase space of H--S--Se at 200~GPa. Large blue and small red circles denote thermodynamically stable and unstable phases, respectively, and grey lines indicate tie lines on the convex hull. The compositional space investigated here with the CE approach is indicated by the yellow line connecting \ce{SeH3} and \ce{SH3}. The structures of the phases \ce{SH2} (\textit{Cmca}) and \ce{Se2H} (\textit{C2/m}) are taken from Refs.~\cite{li_metallization_2014} and \cite{zhang_phase_2015}, respectively. The S and Se sites were substituted in both phases to compute \ce{SeH2} and \ce{S2H}, respectively. The elemental reference phases are S in the $\beta$-Po structure type~\cite{luo_ensuremathbeta-po_1993}, Se in the bcc structure type~\cite{akahama_structural_1993}, and the \textit{C2/c} phase of molecular \ce{H2}~\cite{pickard_structure_2007}.}\label{fig:Gibbs200GPa}
\end{figure}

In addition to the CE calculations, we perform structural searches at 200~GPa using the Minima Hopping Method (MHM)~\cite{goedecker_minima_2004,amsler_crystal_2010} at the three stable compositions.  The MHM implements a reliable algorithm to explore the low-lying portions of the enthalpy landscape given the chemical composition. Consecutive, short molecular dynamics (MD) escape trials to overcome enthalpy barriers are followed by local geometry optimizations. The Bell-Evans-Polanyi principle is exploited by aligning the initial MD velocities along soft-mode directions in order to accelerate the search~\cite{roy_bell-evans-polanyi_2008,sicher_efficient_2011}. In the past, the MHM has been successfully employed to predict or resolve the structure of a wide class of materials, including superconducting materials at high pressures~\cite{amsler_crystal_2012,flores-livas_high-pressure_2012,amsler_novel_2012,huan_thermodynamic_2013,clarke_discovery_2016,clarke_creating_2017,amsler_prediction_2017,amsler_dense_2017,amsler_exploring_2018}. At least two distinct MHM runs are performed at each relevant composition, using both random structures and the ground states from the CE as the initial seeds. We find no structures with lower enthalpies than the ground states predicted through the CE, confirming that we correctly identify the lowest enthalpy structures at the given stoichiometries.

The three ternary phases above are not only stable along the constrained compositions \ce{S_xSe_{1-x}H_{3}} in the phase diagram, but also with respect to all other competing phases in the S--Se--H system.  The complete Gibbs triangle convex hull is shown in Fig.~\ref{fig:Gibbs200GPa}, where the formation enthalpies are encoded in a color plot as a function of composition. The yellow line connecting \ce{SeH3} and \ce{SH3} corresponds to the compositions sampled with the CE. Note that all phases along this line  on the complete, three-dimensional convex hull of the S--Se--H system are thermodynamically stable, as indicated by the blue circles.

\begin{figure}[htb!]
	\centering
	\includegraphics[width=0.8\columnwidth]{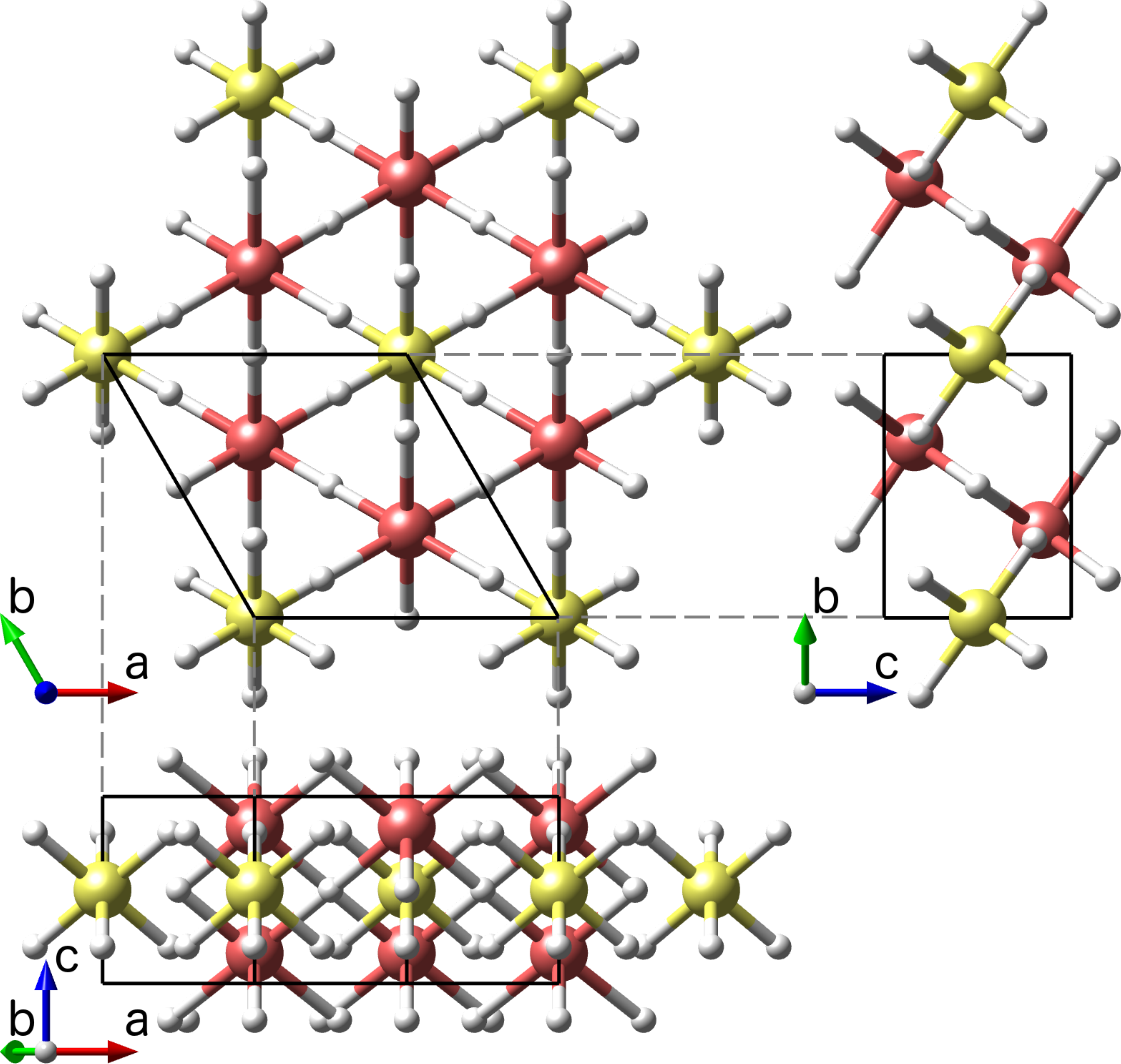}
	\caption{The ground state structure  of \ce{S_{0.$\overline{3}$}Se_{0.$\overline{6}$}H3} with \textit{P$\overline{1}$m1}  symmetry from three different perspectives. The small (white) spheres denote the H atoms, while the large yellow (light) and orange (dark) spheres correspond to the S and Se atoms.}\label{fig:ATAT8}
\end{figure}

Among the phases that constitute the convex hull within the CE, \ce{S_{0.$\overline{3}$}Se_{0.$\overline{6}$}H3} is exceptional due to several reasons. First, we observe the lowest enthalpy among all phases at this composition. Further, in contrast to both \ce{S_{0.2}Se_{0.8}H3} and \ce{S_{0.6}Se_{0.4}H3} which barely touch the convex hull (especially \ce{S_{0.2}Se_{0.8}H3}), \ce{S_{0.$\overline{3}$}Se_{0.$\overline{6}$}H3} also denotes the point inflicting the strongest change in the slope of the hull. Second, the enthalpy gap between the ground state and the next higher enthalpy configuration at that given  composition is especially large, namely 58~meV/f.u. (here, the chemical formula is \ce{SSe2H9}). In fact, this enthalpy gap is the largest among all compositions constituting the convex hull from the CE. These two criteria are strong evidences that \ce{S_{0.$\overline{3}$}Se_{0.$\overline{6}$}H3} is thermodynamically particularly stable.

The corresponding ground state structure of \ce{S_{0.$\overline{3}$}Se_{0.$\overline{6}$}H3} has \textit{P$\overline{1}$m1}  symmetry and is shown in Fig.~\ref{fig:ATAT8}. The view along the $c$-axis shows that the Se atoms form a channel-like geometry, surrounding units of \ce{SH6} at its center. Each S is surrounded by six H atoms at the identical distance of 1.386~\AA\, which form bridges to surrounding Se atoms, S--H$\cdots$Se. Note that this S--H bond length is slightly shorter than in pure \ce{SH3} (1.491~\AA), but is close to the S--H bond in molecular \ce{SH2} (1.336~\AA). On the other hand, the Se atoms are surrounded by H atoms with two distinct bond lengths, namely three with 1.555~\AA\, for the Se--H$\cdots$Se bonds, and three with 1.701~\AA\, for the Se--H$\cdots$S bonds. In comparison, the Se--H bond length in \ce{SeH3} has an intermediate value of 1.573~\AA.

We can explain the particularly high stability of \ce{S_{0.$\overline{3}$}Se_{0.$\overline{6}$}H3} in terms of  the properties of its electronic structure. Both \ce{SH3} and \ce{SeH3} exhibit a rather high density of states (DOS) at the Fermi level, $N_{E_F}$. This high $N_{E_F}$ can be attributed to a van Hove singularity in the DOS very close to the Fermi level, which stems predominantly from the anti-bonding states of the S--H and Se--H interactions, respectively, as we see from a COHP analysis using the Lobster package~\cite{deringer_crystal_2011,dronskowski_crystal_1993,maintz_analytic_2013} (see Supplementary Materials). Such high occupations of states at the Fermi level is electronically unfavorable, and lowering the value of $N_{E_F}$ can lead to a decrease of the band energy, and consequently to a lower formation enthalpy. Providing the additional chemical degree of freedom to the system by allowing the mixing of S with Se allows a rearrangement of the atoms and their bonds to decrease the value of $N_{E_F}$, thereby leading to an improved stability. We observe exactly this behavior in the  \ce{S_xSe_{1-x}H_{3}} system. Tab.~\ref{tab:superconductivity} lists the normalized values of $N_{E_F}$ for all relevant phases, and all ternary compounds exhibit significantly lower  $N_{E_F}$, reaching a minimum of  $N_{E_F}=0.038$~eV$^{-1}$ for \ce{S_{0.$\overline{3}$}Se_{0.$\overline{6}$}H3}.

\begin{figure}[t!]
	\centering
	\includegraphics[width=1.\columnwidth]{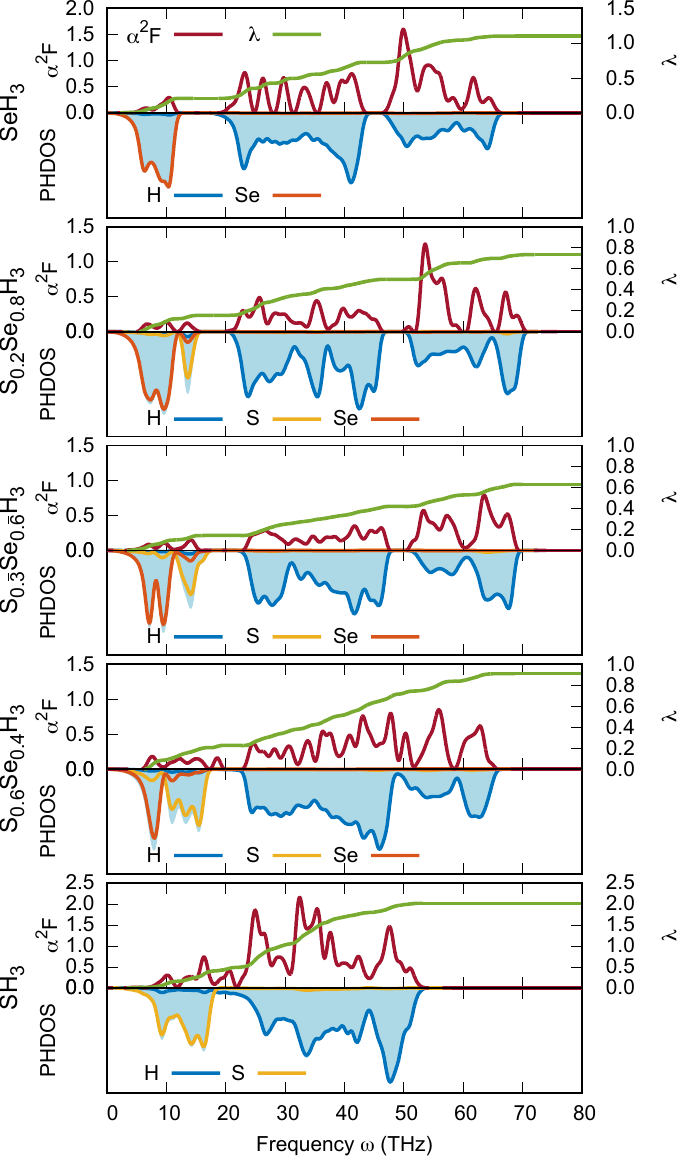}
	\caption{The electron-phonon properties of the thermodynamically table  \ce{S_xSe_{1-x}H_{3}} phases. Each panel corresponds to a specific stoichiometry, with the top part showing the Eliashberg spectral function $\alpha^2F$ together with the integrated electron-phonon coupling constant $\lambda$, and the lower part showing the partial phonon density of states (PHDOS).}\label{fig:elphon}
\end{figure}

This change in the electronic structure also affects the superconducting behavior in \ce{S_xSe_{1-x}H_{3}}. We assess the superconducting properties with the Allan-Dynes modified McMillan's approximation of the Eliashberg equation~\cite{Allen_1975}. The electron-phonon parameters are computed within a linear response framework as implemented in the Quantum Espresso package~\cite{espresso}. The values for the Coulomb pseudopotential $\mu^{*}$ is chosen to be 0.10, which has been shown to give $T_c$ in good agreement with experiments for hydride superconductors, and a Gaussian smearing parameter of $\sigma=0.03$~Ry is used for the double delta integration over the Fermi surface to compute the electron-phonon interactions. We use norm conserving pseudopotentials~\cite{troullier_efficient_1991} and a  plane-wave cutoff energy of 60~Ry, together with dense $k$-point meshes to ensure convergence of the $T_c$ values (see Supplemental Materials for details).

 The values of $\lambda$ and $\omega_{ln}$ for all ground states are listed in Tab.~\ref{tab:superconductivity}, including the superconducting transition temperature estimated within the Allan-Dynes modified McMillan's approximation of the Eliashberg equation~\cite{Allen_1975}. Note that for the composition \ce{S_{0.6}Se_{0.4}H3},  the ground state and the first excited state are very close in enthalpy, merely 6~meV/f.u. apart (i.e., 0.3~meV/atom). Since both phases are essentially degenerate in enthalpy, we report here the superconducting properties of only the one with a smaller unit cell (1~f.u., 20 atoms per cell) to reduce the computational cost. Overall, the superconducting parameters are in agreement with the values found in the literature for the previously reported phases  of \ce{SH3}~\cite{duan_pressure-induced_2014} and the metastable phase \ce{S_{0.5}Se_{0.5}H3}~\cite{liu_effect_2018}. 

\begin{center}
\begin{table}[t!]
\begin{ruledtabular}
\begin{tabular}{l@{}c*{4}{c}}
    Phase                                     & $N_{E_F}$ (eV$^{-1}$)    & $\lambda$  & $\omega_{ln}$ (K)   & $T_c$ (K)\\ \hline
    \ce{SeH3}             &  0.053  & 1.10 & 1379 & 110 \\
    \ce{S_{0.2}Se_{0.8}H3} &  0.039 & 0.73 & 1400 & 54 \\ 
    \ce{S_{0.$\overline{3}$}Se_{0.$\overline{6}$}H3}&  0.038&0.63 & 1448 & 39\\
    \ce{S_{0.5}Se_{0.5}H3}$^*$ & 0.052  & 0.99  & 1421           &  99 \\
    \ce{S_{0.6}Se_{0.4}H3}  & 0.045 & 0.91 & 1414 & 84  \\
    \ce{SH3}                  &  0.055  & 2.02 &1280 & 185
\end{tabular} 
\end{ruledtabular} 
\caption{The electron-phonon properties and the superconducting temperatures of all relevant \ce{S_xSe_{1-x}H_{3}} phases. The density of states at the Fermi level $N_{E_F}$ is given in units of states/cell/eV, normalized per number of (valence) electrons. Note that \ce{S_{0.5}Se_{0.5}H3} (marked with $^*$) is \textit{not} thermodynamically stable, and its properties are only reported here for comparison with Ref.~\cite{liu_effect_2018}
\label{tab:superconductivity}} 
\end{table}
\end{center}

We find that \ce{SeH3} has a lower $T_c$ than \ce{SH3}, a behavior that has been previously attributed to the larger ionic size of Se which leads to a larger electronic screening of the hydrogen vibrations~\cite{flores-livas_high_2016}. However, the change in $T_c$ as a function of composition does not follow a monotonic interpolation between the values of \ce{SeH3} and  \ce{SH3}, as one would rather expect from a virtual crystal approximation~\cite{heil_influence_2015,ge_first-principles_2016}. Instead, we observe a marked minimum in $T_c$ as we move along the S/Se concentration in \ce{S_xSe_{1-x}H_{3}}, with the lowest value for \ce{S_{0.$\overline{3}$}Se_{0.$\overline{6}$}H3}. This trend in $T_c$ is strongly correlated with the value of $N_{E_F}$, which in turn directly affects  $\lambda$. Hence, the property that leads to a high thermodynamic stability is essentially responsible for a reduced superconducting transition temperature.

The detailed features of the Eliashberg spectral function $\alpha^2F$, the integrated electron-phonon coupling constant $\lambda$, and the partial phonon density of states (PHDOS) are shown in Fig.~\ref{fig:elphon}. As expected, none of the phases exhibit imaginary phonons, and are therefore dynamically stable. Note how the phonon spectra are roughly split in three regions: the low-frequency Se vibrations, the intermediate S vibrations, and the high-frequency H vibrations. All three regions contribute to the electron-phonon coupling in all phases. However, \ce{SH3} exhibits a spectral function $\alpha F(\omega)$ with especially strong contributions from all phonons of a rather continuous PHDOS distribution. This unique property of \ce{SH3} does not carry over to the ternary mixtures, contributing as a further factor to their reduced $T_c$ values.


In summary, we study the thermodynamic and superconducting properties in the ternary \ce{S_xSe_{1-x}H_{3}} system. We identify three new thermodynamically stable phases at 200~GPa, namely \ce{S_{0.2}Se_{0.8}H3}, \ce{S_{0.$\overline{3}$}Se_{0.$\overline{6}$}H3}, and \ce{S_{0.6}Se_{0.4}H3}. The particularly high DOS due to a van Hove singularity at the Fermi level of \ce{SH3} and \ce{SeH3}, which strongly contributes to their high superconducting temperature, is significantly reduced for all ternary compounds. We attribute this change of the electronic structure to the additional, chemical degree of freedom that allows for a lowering of $N_{E_F}$. As a consequence, the electron-phonon coupling constant  $\lambda$ is  reduced as well, leading to lower superconducting transition temperatures. Hence, alloying \ce{SH3} with Se might not be a viable route towards new compounds with improved superconducting properties, which essentially disrupts the key factors responsible for its high $T_c$. In fact, similar arguments could be applied to other mixtures with elements Y of the form \ce{S_xY_{1-x}H_{3}}, and the chemical constraint to binary \ce{SH_{3}} is essential for the high $T_c$.

\section{Acknowledgments}\label{sec:ack}
We thank Prof. R. Hoffmann, V.I. Hegde, and M.G. Goesten for valuable expert discussions. We acknowledge the support from the Novartis Universit{\"a}t Basel Excellence Scholarship for Life Sciences and the Swiss
National Science Foundation (project No. P300P2-158407, P300P2-174475). 
The computational resources from the Swiss National Supercomputing Center in Lugano (projects s700 and
s861), the Extreme Science and Engineering Discovery Environment (XSEDE) (which is supported by National Science
Foundation grant number OCI-1053575), the Bridges system at the Pittsburgh Supercomputing Center (PSC) (which is
supported by NSF award number ACI-1445606), the Quest high performance computing facility at Northwestern University,
and the National Energy Research Scientific Computing Center (DOE: DE-AC02-05CH11231), are gratefully acknowledged.


%

\end{document}